# The Galactic Black Hole

Mark Morris,
Department of Physics & Astronomy, University of California, Los Angeles, USA

The black hole at the center of our Milky Way Galaxy – the Galactic Black Hole, or GBH, is a rather modest representative of its class. With a mass of 4 x $10^6$ $M_\odot$, it is well over a thousand times less massive than the most extreme supermassive black holes known to be powering the most luminous quasars. Furthermore, the Galactic Black Hole has a remarkably dim accretion flow, and its luminous energy output is overwhelmed by the dense cluster of bright stars and red giants that surround it, except at radio wavelengths. However, the proximity of the GBH compensates for its restrained activity; being over 100 times closer than the next nearest supermassive black hole in a galactic nucleus, it offers us an unparalleled opportunity to observe its behavior in detail. Consequently, far more observational attention has been paid to the GBH and its entourage of stars and gas than to any other single object outside the solar system.

1. **Historical Emergence of the Galactic Black Hole**

Overwhelming extinction toward the Galactic center (~30 visual magnitudes) dictates that the GBH can only be studied in the radio, infrared, or at X-ray energies above ~ 2 keV. Consequently, the location of the Galactic center wasn't really known until advances in infrared astronomy in the 1960's made it possible to see the Galaxy's central concentration of stars (Becklin & Neugebauer 1968). Radio astronomers had already suspected that the dominant radio source complex in that same direction, Sagittarius A, which had been known since the 1950s, coincides with the Galactic nucleus (Downes & Maxwell 1966).

Radio interferometer observations of Sgr A with improved spatial resolution showed that this 10-pc scale source is composed of both thermal and nonthermal substructures, but it was only in 1974 that sufficiently long interferometer baselines made it possible to discover a bright, nonthermal point source lying amidst the extended emission (Balick & Brown 1974)[1]. This point source – later named Sgr A* by Robert Brown (1982) to emphasize its significance within the Sgr A complex – was found to coincide with the center of the Galaxy's stellar cluster within a few arcseconds (the IRS16 complex, Becklin and Neugebauer 1975); indeed, we now know that it defines the dynamical center of the Galaxy. Sgr A* is unique in the Galaxy because its radio brightness is too large to be attributable to a pulsar or any other known type of stellar object, and its spectral energy distribution was found to be rising with frequency, unlike most nonthermal radio sources. The discovery of Sgr A* therefore aroused considerable interest, an interest compounded by the fact that its existence had been somewhat anticipated by Lynden-Bell and Rees (1971), who followed the then-trending hypothesis, originally advanced by Salpeter (1964) and independently by Zel'dovich (1964), that quasars might be powered by accretion onto supermassive black holes residing in galactic nuclei[2]. Lynden-Bell and Rees posited that the prevailing "black hole formation theory of quasars should be extended downwards to black holes of considerably smaller masses". Among their suggested observable manifestations of such a "small" supermassive black hole was that it could appear as a very compact, bright radio source, and that it would be distinctly variable at infrared wavelengths.

---

[1] See Goss, Brown and Lo (2003) for a detailed account of the discovery of Sgr A*.

So the stage was well set for Sgr A* to be considered a strong candidate as a black hole, and indeed, this point of view was presented in a review by Jan Oort (1977), but the decisive fulfillment of that hypothesis would require a demonstration that the object be endowed with a substantial mass confined to a sufficiently small volume. It was quickly determined from VLBI observations that Sgr A* is very compact (< 200 AU; Lo et al. 1975), but not that it had a large mass.

## 2. The Mass of the Galactic Black Hole

The first inklings of a large mass at the Galactic center came from measurements of the velocity dispersion of gas clouds occupying the central parsec. Measurements of the velocities and velocity dispersions of the 12.8 µm fine-structure line of ionized neon were reported in a series of papers by a group at UC Berkeley (Wollman et al. 1976; Lacy et al. 1979, 1980), who favored a model having a point mass of several million solar masses surrounded by a comparable mass of stars within 1 parsec, but they noted that a model in which all the mass is in the form of stars distributed throughout the central parsec was only slightly less probable. Subsequently, Serabyn and Lacy (1985) observed the same line with higher spatial resolution and were able to identify orbiting streams of gas – the "Western Arc" and the "Northern Arm" – and to fit their velocity distributions with orbits responding to a point mass of $\gtrsim 3.5 \times 10^6$ $M_\odot$ plus a smaller amount of distributed stellar mass. They concluded that the mass distribution is too centrally concentrated to be consistent with a spherical isothermal central cluster (*i.e.*, a cluster having density $\propto r^{-2}$, which is roughly consistent with the observed 2 µm light distribution on larger scales; Becklin and Neugebauer 1968), so they therefore attributed their inferred central point mass to a black hole located, within ~2" accuracy, at the position of the IRS16 complex and Sgr A*.

Stellar velocities were also used to assess the mass distribution, as a check on whether the velocity measurements of gas might be affected by non-gravitational effects, such as jets, magnetic fields, supernova remnant dynamics, etc. McGinn et al. (1989), Sellgren et al. (1990), and Haller et al. (1996) measured the velocity dispersion of stars using the shape of the 2.3 µm CO bandheads in the integrated starlight entering apertures of various sizes. Their preferred model invoked a central point mass of $2 - 2.5 \times 10^6$ $M_\odot$, somewhat smaller than had been deduced from measurements of gas.

However, a much clearer demonstration that a point mass is present came when near-diffraction-limited speckle-imaging observations were undertaken over a period of several years by two groups: Eckart & Genzel (1997, 1998); Genzel et al. (1997, 2000); and by Ghez et al. (1998). The proper motions of ~100 stars located within several arcseconds of Sgr A* were measured, and the velocity dispersion as a function of radius, $r$, was found to closely follow an $r^{-1/2}$ law, which was interpreted to imply the presence of a dark mass of at least $2.6 \times 10^6$ M$_\odot$ confined within 0.1 arcsec of Sgr A*. The proper motion velocities agreed statistically with spectroscopically determined velocities for a Galactic center distance of 8 kpc, so the 3-dimensional velocities of the stars in this cluster of what came to be known as "S" stars surrounding Sgr A* appear to be isotropically distributed.

---

[2] More precisely, the theory invoked by Lynden-Bell and Rees was that black holes are formed when quasars die, and that "each dead quasar disappears down its Schwarzschild throat and is surrounded by stars."

The next step in refining the stellar motions was to measure the nonlinearity of the proper motions: accelerations. Ghez et al. (2000) reported the accelerations of three stars, which provided three projected force vectors that triangulated the location of the point mass to lie within $0.05 \pm 0.04$ arcseconds of Sgr A*. Later, the time baseline of the astrometric measurements became sufficient to fit 3D orbits (Schödel et al. 2002; Ghez et al. 2005). Finally, spectroscopy of the orbiting stars was brought to bear, and the radial velocity information obtained from spectra led to refined S star orbits (Eisenhauer et al. 2005; Paumard et al. 2006; Ghez et al. 2008). The radial velocity measurements of orbiting stars introduced another important parameter into the orbital fits: the distance, $R_o$, to the Galactic center. This distance can be determined geometrically by relating the astrometric measurements of proper motion – an angular velocity on the plane of the sky – to the radial velocity measurements obtained from spectroscopy. As a result, the orbital fits yield both the GBH mass and $R_o$, although these two parameters are inevitably strongly correlated, as has been shown in all the treatments to date (Eisenhauer et al. 2003, 2005; Ghez et al. 2008; Gillessen et al. 2009; Boehle et al. 2016; Gillessen et al. 2017; GRAVITY collaboration 2018, 2019; Do et al. 2019a). Except for Boehle et al. (2016), the analyses have all been focused on a single star, S0-2 (or S2), chosen because it is the brightest S star and has a relatively short period of 16 years. The most recently determined values for the GBH mass and $R_o$ are converging on $4.07 \pm 0.1 \times 10^6$ M$_\odot$, and $8.1 \pm 0.1$ kpc, respectively, with the uncertainties expressed here representing the range needed to accommodate the values of the two reporting groups. The formal statistical uncertainties are far less than this, and indeed the Gravity Collaboration (2019) claims an overall uncertainty in $R_o$ of only 0.3%, but unaccounted systematic uncertainties by one or both groups are probably responsible for the divergence between their values. In any case, the accuracy should continue to improve with more orbital data, and with improved control of systematics, and the determination of $R_o$ should soon be extraordinarily precise.

The GBH mass determined from stellar orbits exceeds that determined earlier from statistical analyses and Jeans modelling of stellar proper motions by ~50% (although the mass determined statistically from radial velocity measurements alone – $3.45 \pm 1.5 \times 10^6$ M$_\odot$ – was in slightly better agreement; Figer et al. 2003). This discrepancy is attributable to the fact that the radial density distribution of the relatively luminous stars that were being used for the analyses with respect to galactocentric radius is flatter than was initially assumed in the statistical treatments (Do et al. 2009b, 2013a, b; Buchholz et al. 2009; Fritz et al. 2016), causing the innermost volume where higher-velocity stars would be present, to be relatively underrepresented in the early analyses. Indeed, the luminous late-type stars in the entourage of the GBH – mostly giants – are distributed in a core-like structure that flattens within 8 - 12" (0.3 – 0.5 pc; Figer et al. 2003, see also Sellgren et al. 1990), rather than in a cusp structure peaking sharply on the GBH that is theoretically predicted for a dynamically relaxed distribution (Bahcall & Wolf 1976). However, Gallego-Cano et al. (2018) have argued that faint stars do appear to form a cusp, so some as-yet-unidentified process might have acted to alter the distribution or the intrinsic luminosity of the giants.

Can we be certain that the implied compact mass is indeed a black hole? Other exotic alternatives have been suggested: a cluster of compact stellar remnants (Maoz et al. 1998), a degenerate fermion ball composed of massive dark particles such as sterile neutrinos (Tsiklauri & Viollier 1998), or a boson ball (Torres et al. 2000), but observations of stellar orbits having periapse passages that come within 50 – 125 AU of the center of mass constrain the volume occupied by the mass to such an extent that such possibilities became clearly inviable (Schödel et al. 2002; Ghez et al. 2005). All would quickly collapse to a general relativistic black hole. Much tighter constraints on the size of the Sgr A* radio source from VLBI radio observations (e.g., Shen et al. 2005) made the implied minimum density of the central mass even far larger. Finally, now that an actual supermassive black hole has made an appearance in the galaxy M87, thanks to the Event Horizon Telescope Collaboration (2019), there is no impediment to assigning black hole status to Sgr A*.

### 3. General Relativity and the Galactic Black Hole

Precision measurements of stellar orbits close to the GBH provide an opportunity to explore General Relativity (GR) in a poorly tested regime. The GR measurements relating to the GBH to date have all been focused on the star S0-2, which is the brightest of the S stars, and which has a relatively short period – 16 years – in its rather eccentric orbit (e = 0.89). The first post-Newtonian test has been the measurement of the gravitational redshift, reported by two groups following the 2018 periapse passage of S0-2 at a distance of 120 AU from the GBH (GRAVITY Collaboration 2018a; Do et al. 2019). At that point, the star was moving at 2.7% the speed of light. Figure 1 shows the projected orbit and the excursions of radial velocity as a function of time from the measurements used by the GRAVITY Collaboration (2018a), including the precision measurements around periapse by the GRAVITY interferometer. The conclusion of both groups was that the data deviate very significantly from a Keplerian orbit, and by an amount that is consistent with the predictions of special and general relativity. Figure 2 shows the deviations from a Keplerian orbit measured with the Keck Telescopes by Do et al. (2019). The deviations are parameterized by the quantity $\Upsilon$, such that the deviation can be expressed as

$$\text{Deviation from Keplerian orbit as a function of time} = \Upsilon \left[ \frac{V(t)^2}{2c} + \frac{GM_{GBH}}{cR(t)} \right], \quad (1)$$

where the first term is the transverse Doppler shift predicted by special relativity, and the second term is the gravitational redshift predicted by GR. $V(t)$ is the 3D velocity of S0-2 at time $t$, $R(t)$ is the 3D distance between Sgr A* and S0-2, and $M_{GBH}$ is the GBH mass. The prediction of classical special and general relativity is that the parameter $\Upsilon$ is exactly 1, and the orbital fitting from both groups is consistent with that value within their $1\sigma$ uncertainties.

The next most challenging GR test using the GBH is to find the Schwarzschild precession of the orbits of nearby stars. The GR prediction is that the periapse of a stellar orbit will undergo a prograde precession. However, that effect will compete with a *retrograde* precession that would be caused by the presence of any significant extended mass distribution (e.g., Rubilar & Eckart 2001). The GRAVITY Collaboration (2020a) recently reported a retrograde shift in the major axis of the orbit of S0-2 of 12 arcminutes per orbital period, which is consistent, within the measurement uncertainties, with the pure GR prediction. That study also concluded that the total amount of extended mass within the orbit of S0-2 must be at least 1000 times less than the mass of the GBH, which places a tight constraint on the abundance of stars and stellar remnants in that volume. Improvements in that limit are sure to be had with continued observational refinement of the orbit of S0-2, and with the addition of more orbits that sample larger and smaller volumes than S0-2. In principle, with accurate measurements of the Schwarzschild precessions of several stars, the radial distribution of extended mass can be determined, and any possible effect that mass may have on the GR test will be eliminated.

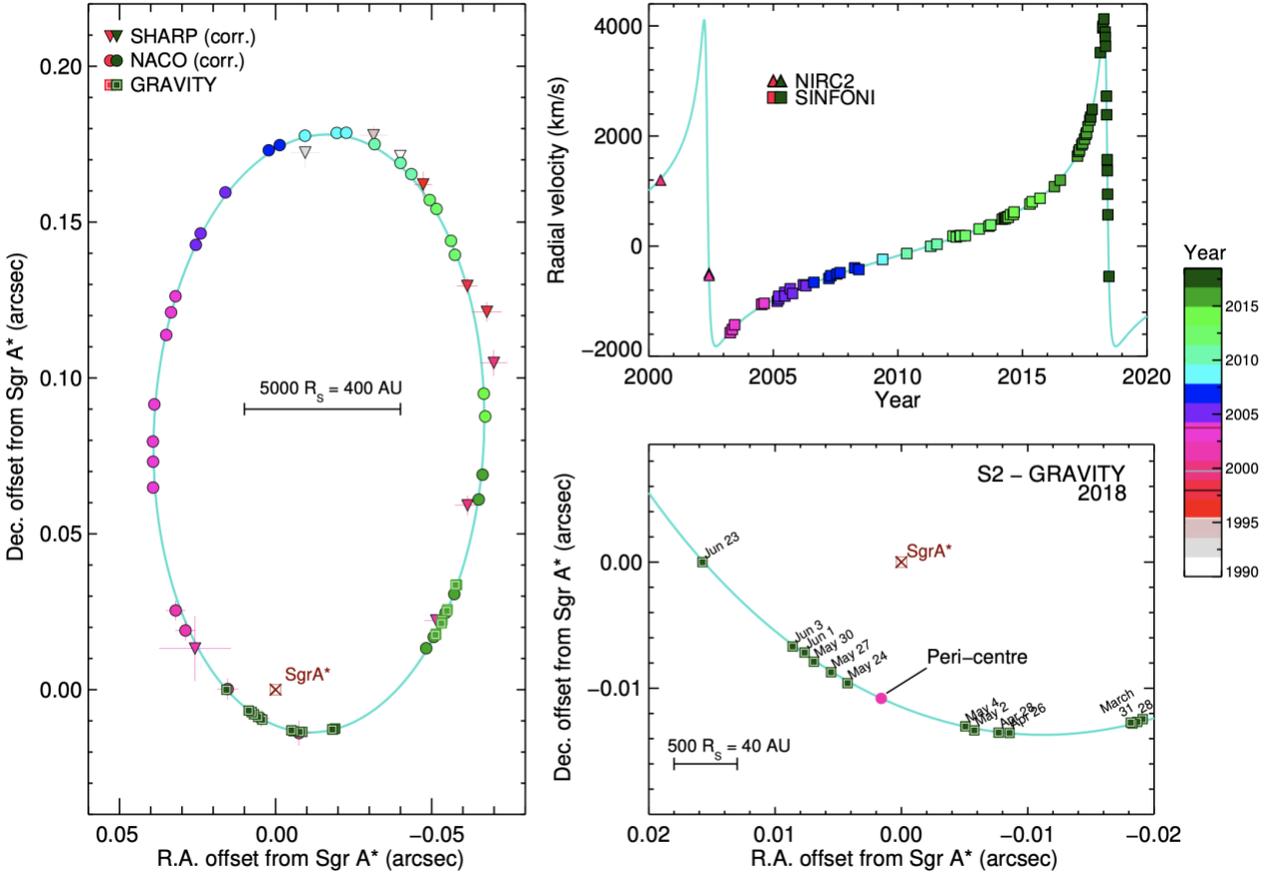

**Figure 1**: Orbit of S0-2 around Sgr A*.  *Left*: orbit projection on the sky.  *Top right*: radial velocity measurements as a function of time.  *Bottom right*: precision positional offset measurements between S0-2 and Sgr A* by the GRAVITY interferometer on the VLT during the passage near periapse.  Figure from Gravity Collaboration (2018a).

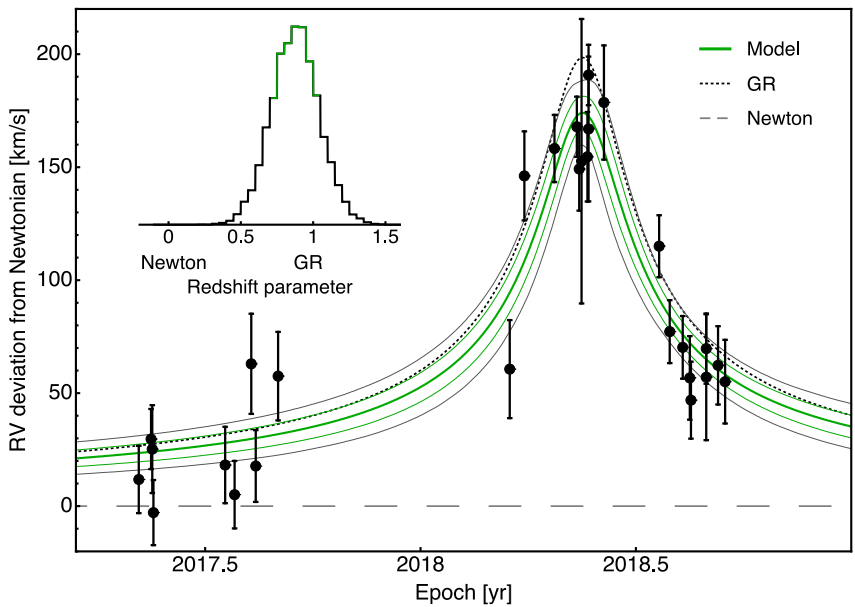

**Figure 2**: Measured deviation of the orbit of S0-2 from Newtonian predictions (from Do et al. 2019).  The green line shows the best-fitting orbit model corresponding to equation (1), and the green and grey shaded regions correspond to the 68% and 95% confidence intervals of the model.  The inset shows the posterior probability distribution for ϒ, which is within 1σ of the predicted relativistic value of 1.

On the more distant horizon, stellar orbits might eventually be used to measure higher-order effects such as the Lense-Thirring precession that accompanies the spin of the black hole, but because that effect falls off rapidly with distance, it would likely require a sample of much more tightly bound stars than we have at present.

## 4. The Accretion Flow: Radiative Signatures of the Galactic Black Hole

The accretion flow onto the GBH emits across the spectrum, though its total luminosity is quite modest, estimated at a mere few times $10^{35}$ $L_\odot$ (Serabyn et al. 1997), which is only ~ $10^{-9}$ - $10^{-10}$ of its Eddington luminosity. Clearly, the accretion rate is quite small, but in addition, the accretion flow is likely to be radiatively inefficient, as discussed below.

Accessible windows onto the spectrum include the radio through submillimeter, the near- and mid-infrared, and X-rays. The spectrum at radio wavelengths longer than 20 cm is not well determined because of confusing bremsstrahlung emission from surrounding gas, coupled with strong interstellar scattering that broadens the source, while the infrared from 10 to 350 µm is so far overwhelmed by the dust emission from the surrounding Sgr A West HII region. Gamma-ray emission has been reported toward the location of the GBH, but with a relatively low spatial resolution that makes it difficult to unambiguously identify the source as the GBH; there are alternatives, including a nearby pulsar wind nebula (Wang et al. 2006; Acero et al. 2010, see Figure 6).

The spectrum of SgrA* has a peak of ~5 x $10^{35}$ ergs s$^{-1}$ in $\nu L_\nu$ at around 350 µm (Serabyn et al. 1997), known as the "submillimeter bump", where the partially polarized emission (e.g., Marrone et al. 2007) is attributable to synchrotron radiation (Figure 3). Early radio results at relatively long radio wavelengths found that the emission from Sgr A* is variable (Brown & Lo 1982), but that variability was appropriately attributed to refractive interstellar scintillation caused by inhomogeneities in the intervening interstellar medium (Zhao et al. 1989). Because of the $\lambda^2$ scaling of refractive scintillation, the intrinsic variability of Sgr A* emerges at higher frequencies; the intrinsic radio emission through the submillimeter bump is variable at the 30 – 40% level on time scales of minutes to hours (e.g., Zhao et al. 2003; Macquart and Bower 2006). The statistical characterization of the millimeter-wave emission was found to be describable in terms of a red-noise power spectrum, in which the power spectral density, $P$, at temporal frequency $f$ is expressible as $P(f) \propto f^{-1}$ (Mauerhan et al. 2005).

4.1 Infrared Emission from Sgr A*

The infrared counterpart to Sgr A* was discovered independently by Genzel et al. (2003) and Ghez et al. (2004), both using new systems that employed adaptive optics on large telescopes for the first time. In both reports, it was clear that the infrared emission is highly variable. Like the millimeter-wave emission, the infrared follows a red-noise power spectrum, although with $P(f) \propto f^{-2.2}$ (Do et al. 2009; Witzel et al. 2018), much steeper than was found for millimeter waves.

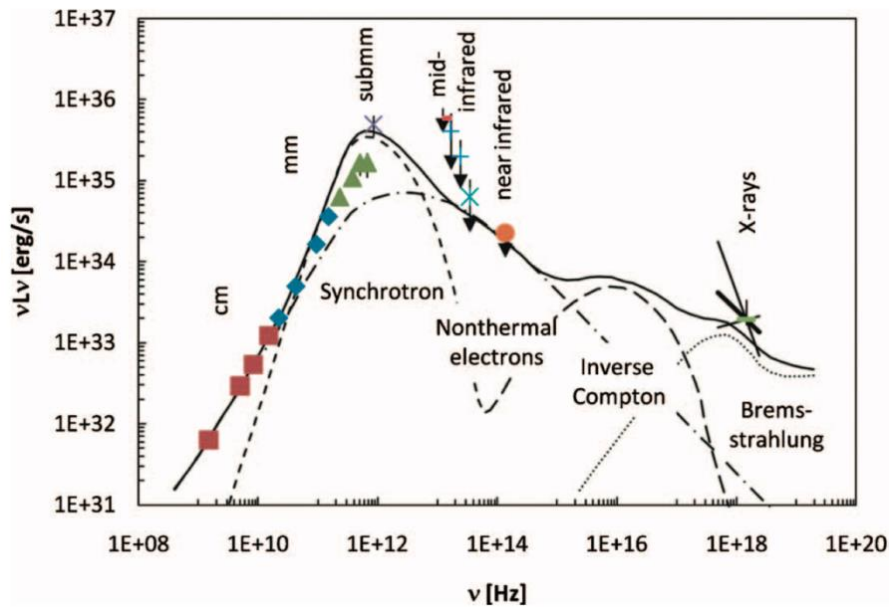

**Figure 3:** Spectral energy distribution of Sgr A* in its steady state, from Genzel et al. (2010). The quiescent X-ray flux level shown here is attributed to thermal Bremsstrahlung from distances out to the Bondi radius, but the X-ray flares, arising from within several Schwarzschild radii (§ 4.2 below), have luminosities rising up to several hundred times that level. A model of the emission from Yuan et al. (2003) is superimposed, showing the various processes that comprise the overall spectral energy distribution. Emission in the mid-and far-infrared is difficult to measure because of the bright thermal dust emission surrounding Sgr A* (Schödel 2011), but has recently been reported by von Fellenberg et al. (2018).

Sgr A* stands out in the near-infrared for being very red compared to the stars around it (Figure 4). Hornstein et al. (2007) found that its intensity rises with increasing wavelength through the near-infrared and they concluded that the spectral index appears constant. However, a later treatment of much more data led Witzel et al. (2018) to conclude that the spectral index depends on the source intensity (see also Ponti et al. 2017 and references therein). The source becomes bluer in brighter states, although the spectrum is always rising with wavelength through the infrared band. In spite of the increasing flux density with wavelength, Sgr A* has not yet been detected reliably at wavelengths greater than 5 µm, largely because in the mid-infrared, thermal dust emission in the complex Galactic center region overwhelms the emission from Sgr A* (Schödel et al. 2011). However, detecting the variable emission from Sgr A* may soon be possible with the James West Space Telescope because of its expected pointing and gain stability.

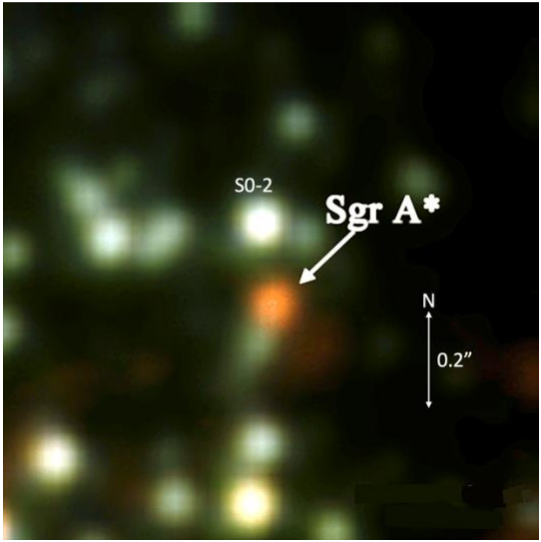

**Figure 4:** 3-color image of the central arcsecond, as observed in July 2005 at a moment when Sgr A* was particularly bright, from Hornstein et al. (2007). Blue: H-band (1.63 µm), green: K' band (2.12 µm), red: L' band (3.78 µm). The relatively bright star to the north of Sgr A* is S0-2, whose orbit has been used to measure the mass of the GBH, and to explore the effects of general relativity. The extended red emission to the southwest of Sgr A* is a dust emission feature associated with the likely stellar source, G1 (Witzel et al. 2017).

The statistical characteristics of the infrared variability of Sgr A* have been investigated on several occasions using data from large telescopes equipped with adaptive optics (Dodds-Eden et al. 2011; Witzel et al. 2012), and by the Hubble Space Telescope, but the most complete characterization to date used eight days of observations at a wavelength of 4.5 µm with the Spitzer Space Telescope between 2013 and 2017 (Figure 5, from Witzel et al. 2018), along with 8 years of $K_s$-band (2.18 µm) data from the VLT and 10 years of $K$'-band (2.12 µm) data from the Keck telescope. As had been found earlier (Do et al. 2009), Sgr A* was shown to be continuously variable, but typically exhibits several prominent maxima per day. Those maxima are often termed "flares," but that designation doesn't capture the continuous and random nature of the variable emission process, which shows variation across a broad range of temporal frequencies. Early reports of a quasi-periodic infrared signal from Sgr A* (Genzel et al. 2003) have not been confirmed; Do et al. (2009) and later, Witzel et al. (2018) found no excess power in their periodogram analyses of their extensive databases at frequencies that could correspond, for example, to the orbital frequency of accreting material near the innermost stable circular orbit (ISCO) around the GBH. This is all the more curious since one might expect to see periodicities from rapidly orbiting gas blobs near the ISCO. Indeed, the GRAVITY Collaboration (2018b) reported orbital motions of blobs near the ISCO having periods of 33 – 65 minutes, which is just outside the ISCO for a Schwarzschild (zero-spin) black hole of 4 million solar masses. Perhaps such blobs do not survive long enough to leave a periodic signal.

In their statistical analysis, Witzel et al. (2018) identified a break frequency in the power spectral density of the infrared emission, indicating a characteristic coherence timescale of $243^{+82}_{-57}$ minutes[3]. This is the time over which the activity displayed in the light curve remains correlated, and can be interpreted as the lifetime of an active region, or the duration of the consequences of an energy release, perhaps by magnetic reconnection (Dexter et al. 2020) or of the sudden introduction of a mass concentration into the accretion flow. A similar characteristic time scale was found for submillimeter light curves by Dexter et al. (2014), although the fractional amplitude variability is less at those wavelengths than in the near-infrared.

---

[3] This verifies and refines an earlier estimation by Meyer et al. (2009).

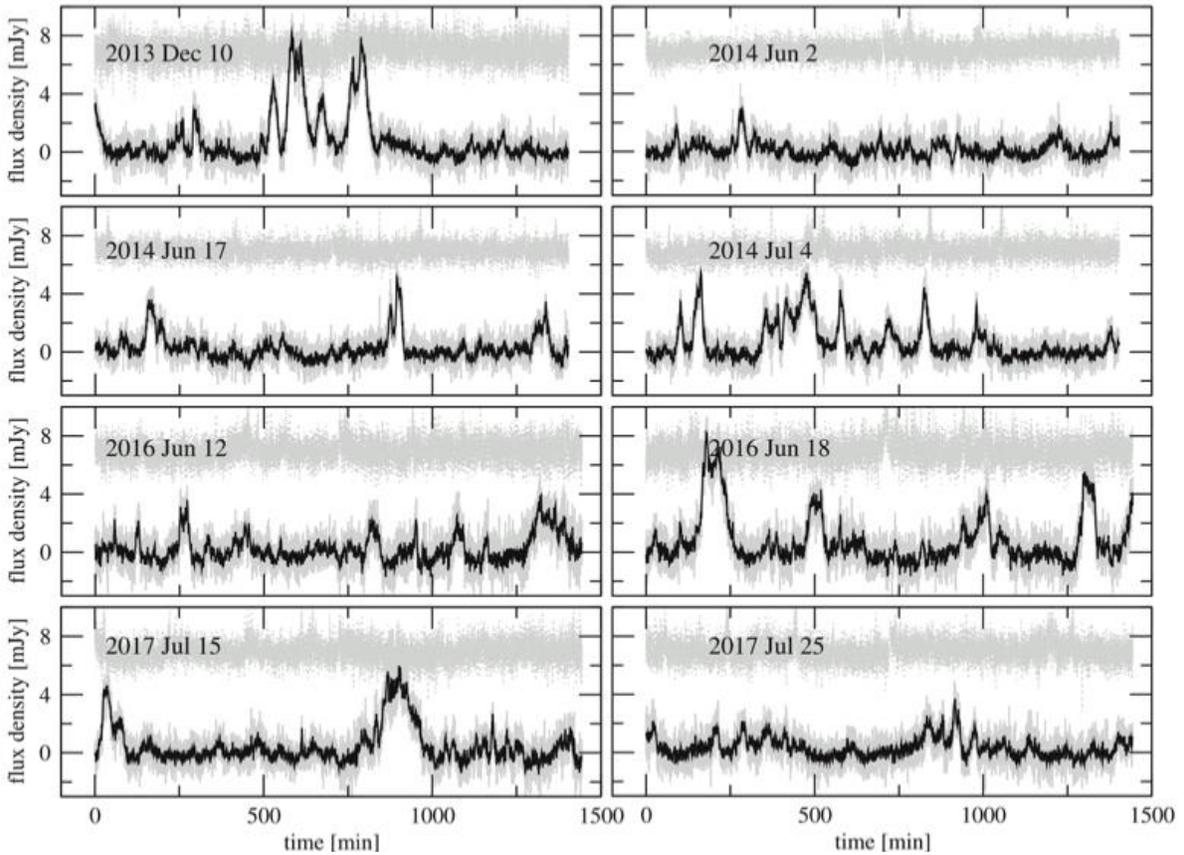

**Figure 5**: 24-hour light curves of Sgr A* at 4.5 μm, as measured by the Spitzer/IRAC camera (Hora et al. 2014; Witzel et al. 2018). In each panel, the gray lines show the flux density for each 6.4 s frame set, and the black lines show the data binned in 1-minute intervals. The light-gray light curves at the top show the flux density of a reference pixel. An additional three 16-hour light curves from 2019, also with IRAC at 4.5 μm, were later published by Boyce et al. (2022).

So far, there is no evidence that the statistical properties of the infrared emission have changed during the time that the infrared emission from Sgr A* has been observed. Following a suggestion that the infrared flux density distribution of Sgr A* switches between two distinct states (Dodds-Eden et al. 2011), Meyer et al. (2014) investigated the data available at that time using a formal methodology – a "Hidden Markov Model," known in economics research as the "regime switching model", and concluded that the emission from Sgr A* is sufficiently described by a stochastic process in a single intrinsic state. Later, Chen et al. (2019) extended the time frame over which Sgr A* had been detected by reanalyzing data that had been obtained prior to 2005 with a speckle imaging technique. Using a speckle holography analysis, Chen et al. extended the depth of the images from that time period by a few magnitudes compared to the previously used shift-and-add technique, and they were able to extract a number of detections of Sgr A* between 1998 and 2005. The resulting source fluxes were found to be consistent with the flux density distributions obtained during the later (≥ 2005) adaptive optics epochs, indicating that no obvious state changes had taken place in the earlier period, and that the variability characteristics of Sgr A* had been unchanged over 22 years. This conclusion is salient when one considers the evidence from X-rays (discussed below) that major outbursts of Sgr A* have occurred on time scales of hundreds of years.

However, a hint that anomalous outbursts can occasionally happen occurred in 2019, when a brightness excursion that was twice as bright as anything previously observed was captured at the Keck Observatory by Do et al. (2019b). Those authors also presented light curves for other unusually bright events that occurred within weeks of the giant flare, and concluded that the probability of the flux levels in the ensemble of their light curves is less than 0.05%. This raises the possibility that Sgr A* was going through a phase of unusual activity in 2019, and Do et al. considered the possibility that such activity might have been the delayed result of enhanced accretion caused by the 2018 periapse passage of the bright S star, S0-2, or even the earlier periapse passage of the dusty, extended G2 object in 2014, which underwent a clear tidal interaction with the GBH. The latter possibility was investigated further by Murchikova (2021), who concluded that material pulled off of G2 by the tidal interaction could rain down later on the GBH and produce enhanced accretion luminosity. Indeed, Kawashima et al. (2017) had predicted a radio and infrared brightening in about 2020 as a result of the G2 passage. The ensemble of 2019 data assembled by the GRAVITY Collaboration (2020b) provides additional evidence that Sgr A* was unusually active during 2019; it shows several near-IR intensity peaks that are markedly stronger than anything that had been seen in previous years, even though the median intensity was unchanged from year to year.

The infrared emission from Sgr A* displays strong linear polarization (Eckart et al. 2006, 2008; Trippe et al. 2007; Zamaninasab et al. 2010; Shahzamanian et al. 2015), which confirms that the infrared emission, like the radio, is optically thin synchrotron emission. In addition, the variability of the polarization adds a new dimension to the study of the infrared emission. Not only does the polarized signal vary largely in concert with the total intensity, but the polarized light curve is also more complex, consisting of "subflares" as the polarization fraction, typically ranging between 10% and 30%, undergoes marked changes through the total intensity maxima. Furthermore, in the course of an infrared flare, the orientation of the polarized E-vector often shows large and rapid changes on time scales of several minutes. These "swings" in the polarization angle have inspired consideration of two classes of models in the references listed above. The most emphasized model involves a "hot spot" orbiting near the innermost stable circular orbit (ISCO), perhaps as part of a transitory disk. The hot spot could be a density enhancement of emitting matter or a location in the accretion flow at which energy injection has taken place as a result of some local instability. The swings in polarization angle are then attributed to the changing projection of the ambient magnetic field within or around the hot spot as it orbits. This appeal of this model has been considerably enhanced by the direct observation of orbiting hot spots near the ISCO by the GRAVITY Collaboration (2018b), accompanied by polarization angle changes that varied continuously with orbital phase and with about the same 45±15-minute period.

The second model for the polarization variations invokes emission from a jet or a collimated wind. The polarization angle from an outgoing plasma blob in a helically structured jet, for example, could change with location in the jet as the projected direction of the magnetic field changes along the helix. In general, however, the polarization angle should reflect the jet direction, and thereby vary around a particular angle. Eckart et al. (2006) discuss potential relationships of their observed polarization angles to possible outflow or jet signatures in X-ray and infrared images, and Shahzamanian et al. (2015) identify a preferred polarization angle – 13 ± 15 degrees – and suggested that it could be linked either to an outflow direction, be it that of a jet or a wind, or to the orientation of a temporary accretion disk. Continuing observations with GRAVITY will be of great value in refining the conclusions that might be drawn from polarization measurements.

The submillimeter emission from Sgr A* also displays a rather strong linear polarization, and like the near-infrared emission, it varies on similar time scales, with the polarization angle undergoing rapid occasional swings (Marrone et al. 2006; Wielgus et al. 2022b).

4.2 Variable X-ray Emission from Sgr A*

X-ray emission from Sgr A*, discovered by Baganoff et al. (2001, 2003), consists of two components: a relatively weak, measurably extended quiescent component that has been invariant for at least 14 years (Yuan and Wang 2016), and an unresolved component consisting of occasional dramatic "flares" that rise above the quiescent component by factors up to several hundred (Wang et al. 2013). The quiescent component (Figure 6) has been attributed to the virialization of the accreting gas at the Bondi radius, where it shocks and thermalizes at a temperature of ~$10^7$ K (e.g.,

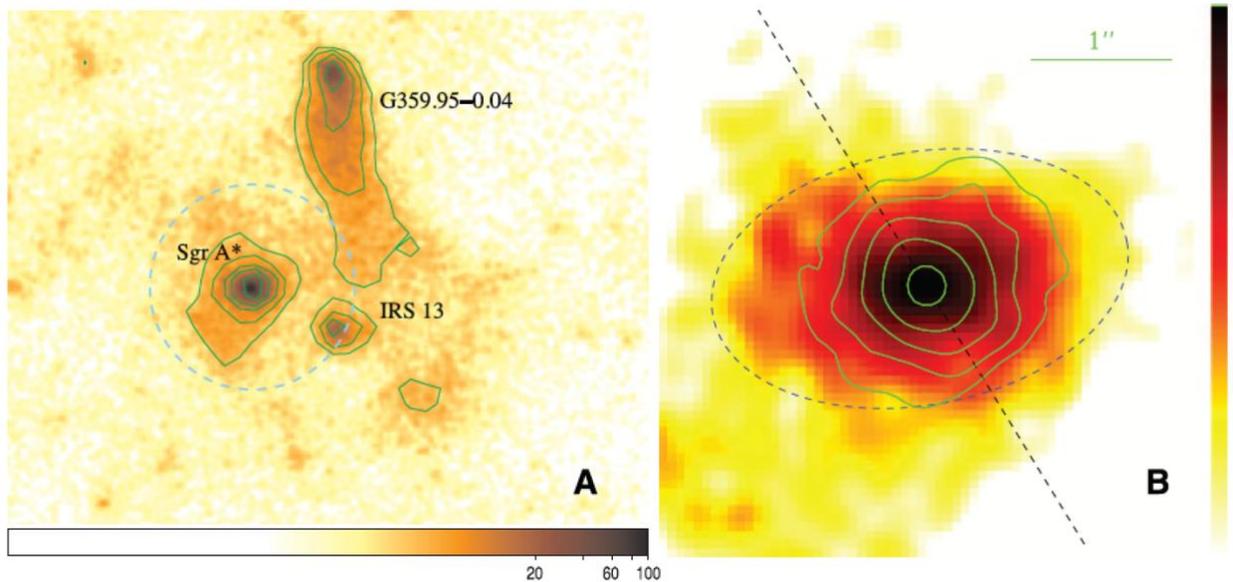

**Figure 6**: X-ray image of Sgr A* from 1 to 9 keV, as measured by the Chandra X-Ray Observatory (from Wang et al. 2013; see also Shcherbakov & Baganoff 2010). *Left*: image of the quiescent emission, with flare intervals removed from the integration. The dashed circle surrounding SgrA*, with a radius of 4 arcseconds, corresponds to the estimated Bondi capture radius. Note the nearby pulsar wind nebula, G359.95-0.04, which is a candidate source of the gamma-ray emission from this region. *Right*: a magnified version of the image at left. The contours show the unresolved flare component. The dashed line indicates the orientation of the Galactic plane, and the dashed ellipse depicts the projection and orientation of the central disk of massive young stars, extending well beyond the ellipse.

Corrales et al. 2020). The Bondi radius occurs at ~ $GM/c_s^2$, for sound speed $c_s$, corresponding to ~0.1 pc (Quataert 2002; Baganoff et al. 2003; Yuan 2011; Wang et al. 2013).

The detectable X-ray flares, which occur about once per day on average, have durations of 0.5 to several hours and often show considerable temporal substructure (Baganoff et al. 2001; Nowak et al. 2012; Ponti et al. 2017; Haggard et al. 2019), including an occasional precursor (Figure 7). The timescales indicate that the flares arise from a region having a size of several gravitational radii, and those regions are probably quite close to the GBH. However, the flare mechanism has not been reliably identified; there remain several possibilities: magnetic reconnection events (Ball et al. 2016, 2020), hydrodynamic or magnetohydrodynamic instabilities in the inner accretion disk (Tagger & Melia 2006), a strongly enhanced accretion rate associated with an inspiraling lump of material, shocks caused by the impact of a jet on material in the accretion flow, gravitational lensing of hot spots in the accreting plasma or even tidal disruptions of comets, asteroids or planets (Zubovas et al. 2012).

4.3 Emission Mechanisms, from Radio to X-rays and Frequency-Dependent Time Lags

There has been much discussion in the literature about the emission mechanism powering the X-ray flares, and the debate continues, primarily between synchrotron emission and synchrotron self-Compton (SSC) emission. The submillimeter and near-infrared emission are readily explained as synchrotron emission, and many have regarded the X-ray flares as a continuation of the direct synchrotron emission causing the infrared emission (e.g., Dodds-Eden et al. 2010; Barriere et al. 2014; Dibi et al. 2014, 2016; Ponti et al. 2017; Zhang et al. 2017; Gravity Collaboration 2021), albeit with a variable cooling break in the electron energy distribution that is needed to account for the absence of detectable X-ray counterparts to most bright infrared maxima, while all sufficiently well observed X-ray flares are coincident with a near-infrared (NIR) maximum (Fazio et al. 2018), indicating that the emission mechanisms for NIR and X-rays are inherently linked.

The SSC model rests on the supposition that the electrons responsible for producing the low-energy photons by synchrotron radiation can also elevate the energies of those same photons to the X-ray regime via inverse Compton scattering. This model was forwarded by Falcke & Markoff (2000) and Markoff et al. (2001) in the context of their jet model for the radio and X-ray spectrum of Sgr A*. Many later authors have investigated the SSC model for emission from the accretion flow (Liu et al. 2006a, 2006b; Marrone et al. 2008; Trap et al. 2011; Eckart et al. 2012; Ma et al. 2019).

The relative merits of the synchrotron and SSC models have recently been summarized by Boyce et al. (2019), who analyzed all existing measurements of simultaneous observations of Sgr A* in the NIR and X-ray to determine whether there is a significant time lag between the emission peaks in those two wavelength ranges. They found that X-ray peaks may lead the NIR peaks by 10 - 20 minutes, but only with 68% confidence, and that both synchrotron and SSC emission processes are consistent with such a time lag. Additional cases of simultaneous X-ray/NIR observations are needed to obtain a more robust result, although it is quite challenging to arrange for simultaneous observations of an X-ray satellite and a major ground-based telescope equipped with adaptive optics. The question of whether submm maxima can also be associated with NIR or X-ray peaks with a time lag is complicated by the fact that the time lag could become comparable to the range of typical intervals between maxima in the light curves. Tentative NIR-submm time lags from 20 to 200 minutes were listed by Morris et al. (2012), but Fazio et al. (2019) presented examples of simultaneous light curves that showed no such obvious trends. A 2019 flare observed simultaneously in the near-IR, X-ray, and submm was analyzed by multiple groups (Gravity Collaboration 2021; Michail et al. 2021; Boyce et al. 2022), who found that the delay between the X-ray and near-IR peaks was less than a few minutes, but the submillimeter appeared to be delayed by about 30 minutes (although the submm maximum was not actually observed, rendering that conclusion rather uncertain). Unfortunately, what is needed to gather good statistics on the important question of the submm time delay is simultaneous light curves with durations exceeding what is usually feasible from a ground-based submm observatory.

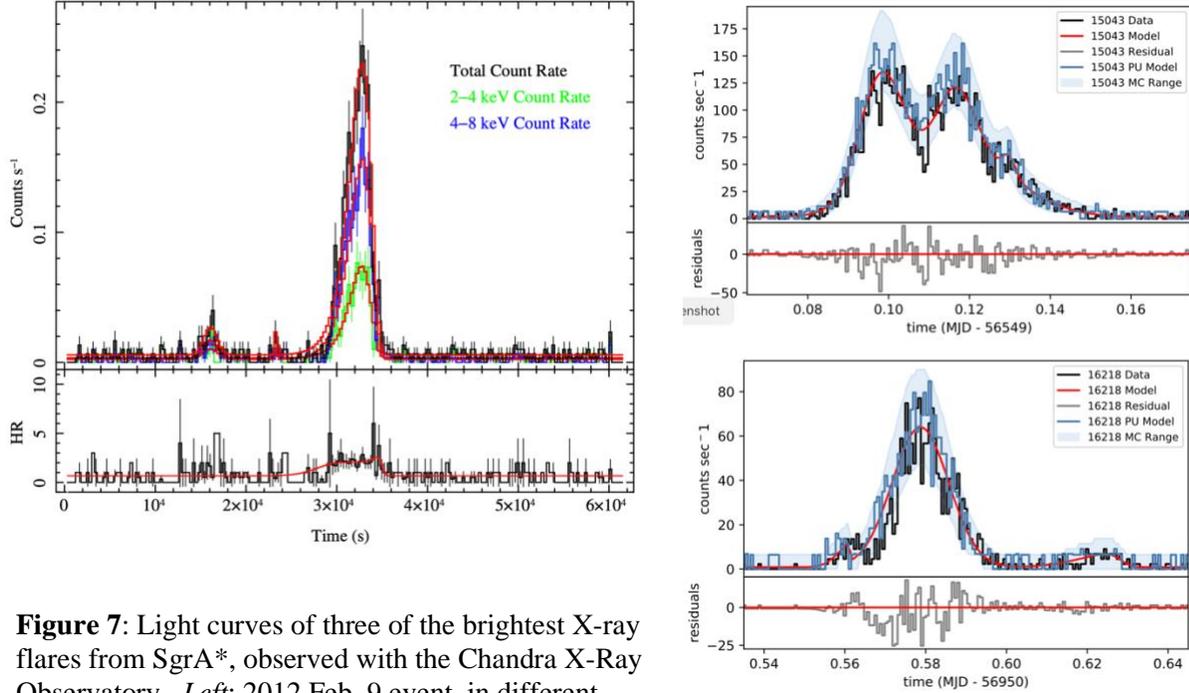

**Figure 7**: Light curves of three of the brightest X-ray flares from SgrA*, observed with the Chandra X-Ray Observatory. *Left*: 2012 Feb. 9 event, in different bands, and the hardness ratio (HR) at bottom (from Nowak et al 2012). *Right*: 2013 Sept. 14 and 2014 Oct. 20 flare profiles and model fits, plus residuals, from Haggard et al. (2019).

The idea of an increasing time lag with increasing wavelength in the emissions from Sgr A* was first suggested by Yusef-Zadeh et al. (2006), who compared light curves at two radio frequencies (22 and 43 GHz), finding a 20 to 40-minute lag between the two frequencies. They interpreted their observation in terms of an expanding plasma blob, invoking a model originally suggested by van der Laan (1966). A similar result using much more extensive observations was obtained by Brinkerink et al. (2015), who observed at 7 frequencies from 19 to 100 GHz and found an increasing time lag with decreasing frequency.

With the expanding plasma blob paradigm, any significant time lags could be combined with source size measurements to estimate the expansion speed of the emitting region. The apparent source size, however, is broadened by interstellar scattering, which closely follows the theoretically expected wavelength-squared law (Shen et al. 2005; Bower et al. 2006, and references therein). Subsequently, Falcke et al. (2009) and later, Johnson et al. (2018), used VLBI measurements taken at several wavelengths between 1.3 and 13 mm, and from a variety of telescopes, to refine the relation for the major- and minor-axis diameters of the elliptical scatter-broadened disk of Sgr A*:

$$\phi_{scattering} = (1.38 \pm 0.013) \times (\lambda_{cm})^2 \text{ (major) and } (0.703 \pm 0.013) \times (\lambda_{cm})^2 \text{ (minor) mas.}$$

Johnson et al. also found that SgrA* has a nearly circular *intrinsic* structure, with a diameter roughly proportional to wavelength between 1.3 and 13 mm: $\phi_{Sgr\ A*} = 0.4$ mas $\times \lambda_{cm}$, which corresponds to 3.2AU $\times \lambda_{cm}$, or $39\lambda_{cm}$ Schwarzschild radii ($R_s$). This stands in contrast to the 7-mm observations of Bower et al. (2014), who reported a rather elongated intrinsic source shape of 35.4 x 12.6 $R_s$. Also, Lu et al. (2018), observing at 1.3 mm, reported evidence for compact intrinsic source structure on scales of ~3 $R_s$.

From the reported size distribution and their time lag measurements, Brinkerink et al. (2015) were able to constrain the expansion velocity of the emitting medium, and they concluded that the radio emission arises in a moderately relativistic, collimated outflow. It is not clear that significant time lags can be found at frequencies at or above 100 GHz; Miyazaki et al. (2013) found no time lag between 90 and 102 GHz. Furthermore, radio emission at millimeter wavelengths and longer likely arises at radii larger than the radii at which the variable infrared and X-ray emission is produced (e.g., Witzel et al. 2021), so that must be accounted for when extending time-lag arguments across the full electromagnetic spectrum. The radio emission might not be physically linked to the shorter-wavelength emission.

In a recent study, Witzel et al. (2021) examined the variability characteristics of Sgr A* across the accessible electromagnetic spectrum from submillimeter to X-rays. Those authors engendered a model in which the correlations between NIR and X-ray emission are attributed to two strictly correlated stochastic processes, one dominating the emission in the NIR, and the other, having less power on short timescales, accounting for the X-ray emission. The coupling between X-rays and NIR indicated by the occurrence of a NIR peak coincident with every X-ray flare evokes a model in which the NIR emission arises in a self-absorbed synchrotron volume having a high-frequency exponential cooling cutoff within or just above NIR frequencies. The same volume generates SSC emission that occasionally contributes to the NIR flux, but is dominant at higher frequencies, accounting entirely for the X-ray emission. A simulation of the resulting spectral energy distribution is shown in Figure 8 for two chosen moments, one which produces a detectable X-ray flare (bottom) and one in which only the NIR is detectable (top). In this model, some random energy injection process causes an enhancement in the electron energy distribution that translates into enhanced synchrotron emission from the submm to the NIR, but also into a higher cutoff in the electron energy distribution, and consequently a higher cutoff frequency in the synchrotron spectrum. As a result, the X-ray flux, while not strictly linked to the NIR flux, varies over a much broader intensity range than the NIR flux, and most of the NIR excursions are accompanied by X-ray flares whose fluence is undetectable because it is below the constant quiescent level arising much further out near the Bondi radius. This is consistent with the X-ray flare luminosity distribution extracted from Chandra data by Neilsen et al. (2013 and 2015): dN/dL $\propto$ L$^{-1.9}$, which implies that ~10% of the quiescent flux is owed to weak, individually undetectable X-ray flares.

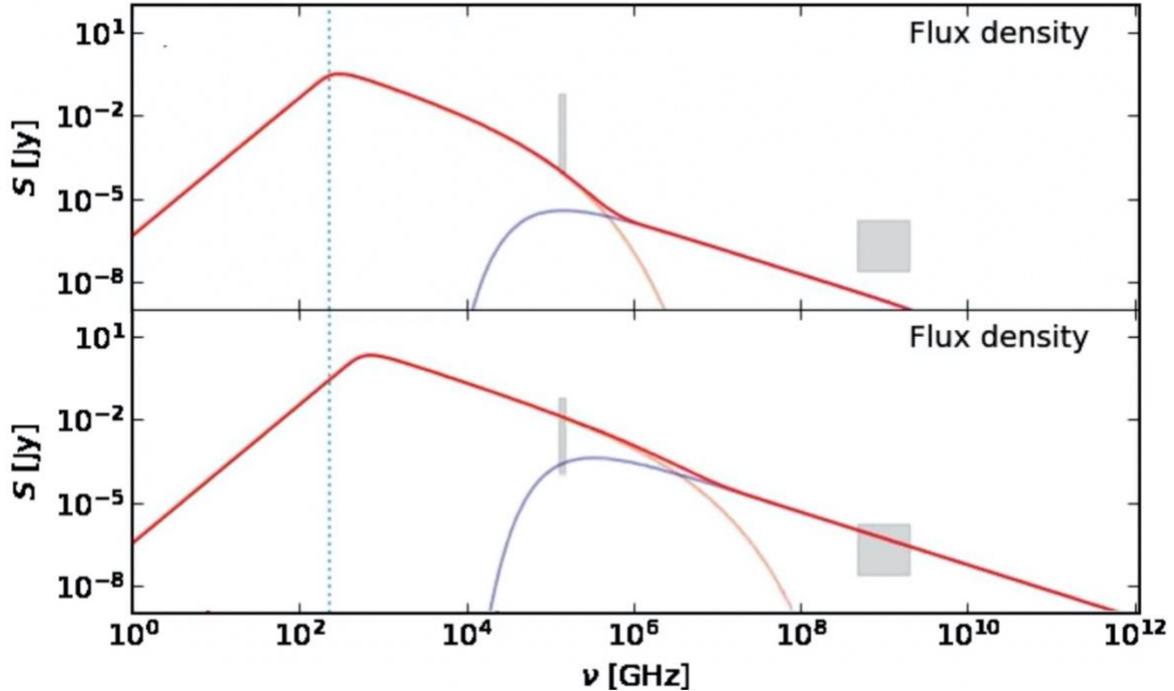

**Figure 8**: Snapshots of the animation of the spectral energy distribution of Sgr A* at two different times, from Witzel et al. (2021). Orange lines show the synchrotron emission, blue lines the SSC emission, and red lines show their sum. The vertical dashed line marks the frequency of 230 GHz, which is near the turnover where the synchrotron emission becomes optically thick. The grey shaded zones show the NIR and 2-8 keV X-ray bands where observations have been made. Their vertical extent corresponds to the typical range of observed NIR variability and the typical range of peak 2-8 keV flux densities above the detection limit. Non-variable radio emission at frequencies below the submillimeter bump located just above 230 GHz arises at larger radii than the variable emission, and is not depicted here, although it dominates the observed flux density.

4.4 Long-term variations in the accretion luminosity of Sgr A*

Evidence has grown over the past decade that multiple powerful pulses of relatively hard X-rays are presently propagating across the central molecular zone of the Galaxy, and the consensus is that those outgoing X-ray fronts originated in an energetic accretion event from the GBH within the past few hundred years. Observers using the Chandra, XMM-Newton, Suzaku and NuSTAR X-ray satellites have followed multiple "light echos" consisting of a combination of fluorescent X-ray line emission and scattered hard X-rays, and that echo has been observed to move at apparent relativistic speeds (even superluminal speeds) in the general direction away from Sgr A* (Muno et al. 2007; Inui et al. 2009; Ponti et al. 2010; Capelli et al. 2012; Nobukawa et al. 2012; Clavel et al. 2013; Zhang et al. 2015; Chuard et al. 2018; Terrier et al. 2010, 2018). The most prominent line is the 6.4 keV Kα line of neutral or singly ionized iron, which results from the photo-ejection of the most tightly bound electron by an X-ray with energy exceeding the 7.1 keV ionization edge of iron. In the ensuing radiative cascade, K lines are emitted, and the Kα line is the most intense.

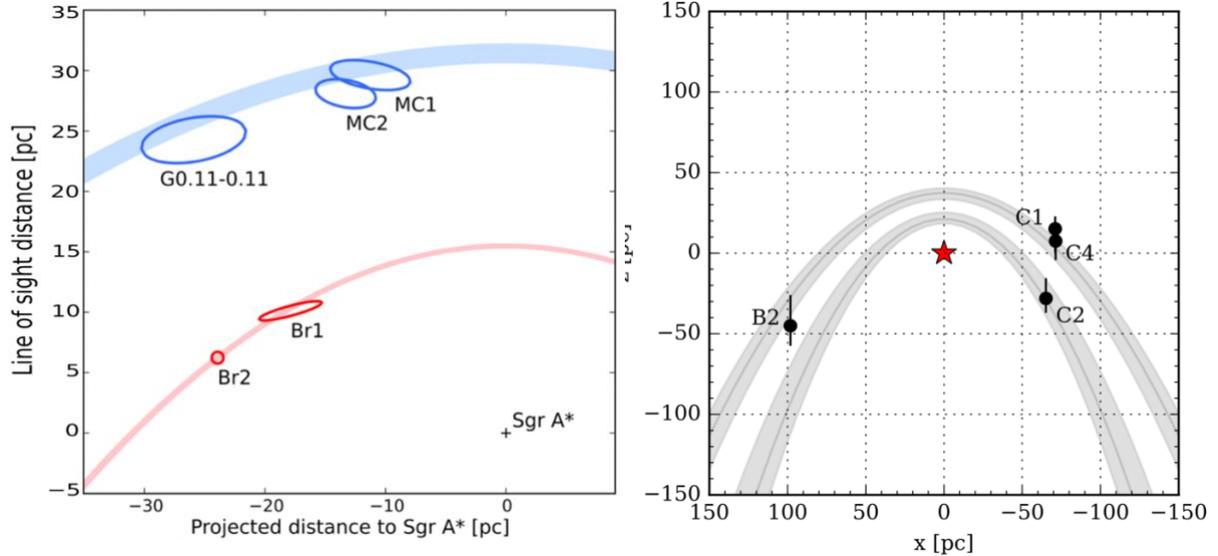

**Figure 9**: Inferred present loci of molecular clouds recently experiencing the intrusion of the hard X-ray fronts presumed to have originated from Sgr A*. *Left*: clouds observed by Clavel et al. (2014). The parabolic curves show the presently observable regions of the planar central molecular zone and the durations of the flashes from two separate events, one launched ~100 years ago (red), and the other launched ~200 years ago (blue. *Right*: same, for a different set of clouds observed by Chuard et al. (2018) near Sgr C, plus the Sgr B2 cloud, in which the fluorescent iron line emission has recently subsided, indicating that a ~10-year X-ray front has largely passed through it (Terrier et al. 2010, 2018). Sgr A* is at the origin. The X-ray fronts in the two studies do not necessarily correspond to the same events, but their inferred durations more or less agree with each other.

This line, and the scattered flux of hard X-rays, are produced in a number of molecular clouds in the central molecular zone as the flash of hard X-rays passes through them (Figure 9).

The fluorescent iron line emission phenomenon was anticipated by Sunyaev et al. (1993) as a means to constrain the occurrence of luminous X-ray outbursts during the past several hundred years. The subsequent detection of bright 6.4 keV line emission toward two molecular complexes in the central molecular zone by Koyama et al. (1996) led those authors to suggest that the Galactic center could have undergone an outburst as strong as $2 \times 10^{39}$ ergs s$^{-1}$. A similar luminosity has been inferred by most later studies. Even higher luminosities are likely implied by the fluorescent emission from the distant molecular cloud, Sgr D (c.f., Terrier et al. 2018), if Sgr A* is indeed the illuminating source.

The duration of the X-ray outbursts from Sgr A* can be constrained by determining how long it takes for the intensity of the 6.4 keV line from a cloud to rise or fall substantially. Of course, that time is affected by the cloud geometry and its relation to the direction of the illuminating source. Taking this into account, Clavel et al. (2012) inferred the presence of at least two events, one lasting ~10 years and the other about 2 years, from observations of clouds at positive Galactic longitudes. Similarly, Chuard et al. (2018), using observations of the Sgr C complex on the negative-longitude side of Sgr A*, inferred from their long-term light curves that there were two illuminating events, one lasting 10 years or longer and the other lasting about 2 years. Figure 9 schematically illustrates, from a terrestrial perspective, the present observable locations of the X-ray fronts and the clouds with which they are interacting.

The occurrence date of the X-ray flashes can be determined if the line-of-sight distance to the fluorescing/scattering clouds can be determined (e.g., Churazov et al. 2017). This is possible because the cross-section for scattering of hard X-rays is dependent on their energy and on the angle subtended at the scattering cloud between the illuminating source (presumed to be Sgr A*) and the line of sight to the observer. Furthermore, that angle, along with the column density of the cloud, also affect the strength of the Fe Kα line and the depth of the iron edge (Walls et al. 2016). The shape of the spectrum and the relative 6.4 keV line flux can therefore be used to infer the scattering angle. The metallicity of the scattering medium must also be taken into account (Chuard et al. 2018). Current estimations of the time of the flashes based on the Sgr B2, Sgr A and Sgr C clouds range between 110 and 250 years ago (Churazov et al. 2017; Chuard et al. 2018), and the variable emission from the more distant Sgr D cloud could imply an even earlier event between 300 and 1100 years ago (Terrier et al. 2018).

The multiplicity of these X-ray flashes having a luminosity on the order of 5 orders of magnitude greater than anything observed over the past 22 years of X-ray monitoring of Sgr A*, and their occurrence in the relatively recent past, make it all the more remarkable that Sgr A* has maintained such a statistically constant emission state during the past quarter century (Chen et al. 2019; Ponti et al. 2015b), with the possible minor exception of the excess activity observed in 2019 (Do et al. 2019b; Gravity Collaboration 2020b).

The occurrence of these hundred-year events could be provoked by accretion of a particularly large lump of gas from the rather inhomogeneous interstellar entourage of Sgr A*, by minor tidal disruption events of small planetary-size bodies, or by tidal stripping of the envelopes of giant stars or G objects that pass perilously close to Sgr A* as they move through their orbital periapse. Tidal disruption events involving stars should occur every $10^4$ - $10^5$ years (*e.g.*, Phinney 1989), and would be much more energetic than the hundred-year events. On even longer time scales, Sgr A* has clearly undergone extremely active stages. A detailed summary of the past activity of Sgr A* has been published by Ponti et al. (2013).

## 5. Dynamics of the Accretion Flow

The source of matter feeding the accretion flow appears to be the winds from the surrounding cluster of young, massive stars (Coker & Melia 1997; Quataert et al. 1999; Quataert 2004; Cuadra et al. 2008; Ressler et al. 2018, 2020a,b; Calderon et al. 2020), which give rise to an accretion rate through the Bondi radius of ~ $10^{-5}$ $M_\odot$ yr$^{-1}$. Because there are numerous moving wind sources of varying strength and velocity, the accretion flow is necessarily quite chaotic and time-variable (Figure 10). Observationally, no concentrations of dust and gas have been identified within about 0.04 pc of SgrA*[4], except for the accretion flow itself and except for matter contributed by the occasional orbital passage of objects from which material can be tidally removed, such as the mysterious G objects, which could be compact dust clouds or distended stellar envelopes (Gillessen et al. 2012; Witzel et al. 2014, 2017; Pfuhl et al. 2015; Ciurlo et al. 2020).

---

[4] There are IR-emitting dust clouds superimposed on or near Sgr A* (e.g., Peissker et al. 2020), but their limited proper motions and small internal velocity gradients imply that they are well displaced from the black hole along the line of sight, and therefore unlikely to be contributing to the accretion flow.

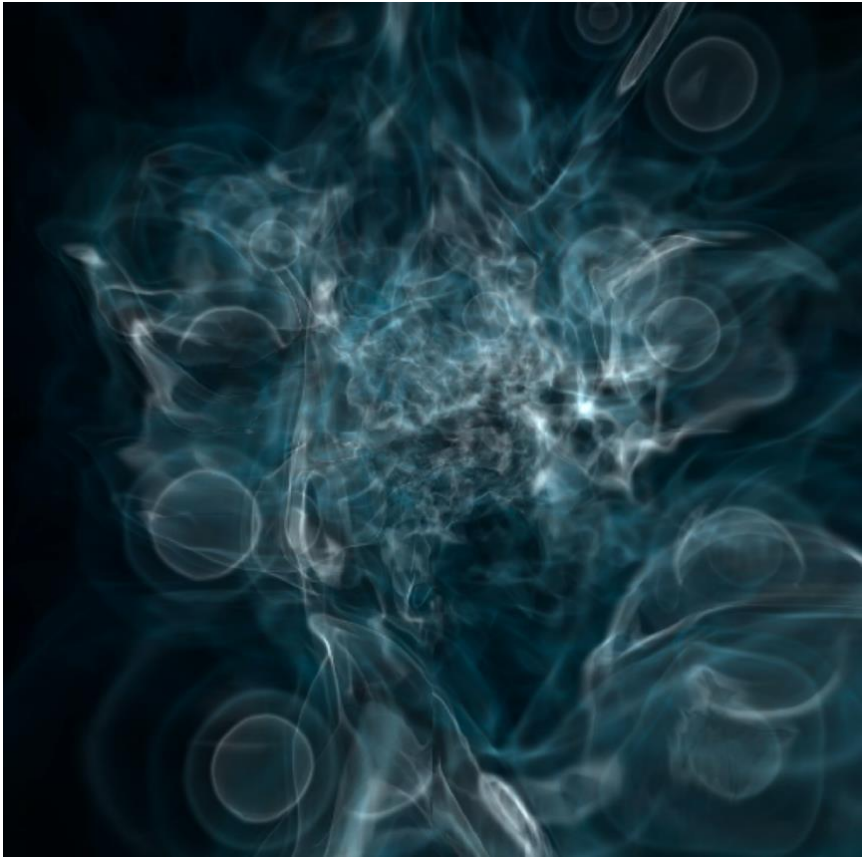

**Figure 10:** Simulation of the accretion flow onto the GBH (centered), and the stellar winds feeding that flow (from Ressler et al. 2018). The panel is 0.5 pc on a side. The stellar wind sources appear as circular rings that form bow shocks as their sources orbit the GBH.

Inside the Bondi radius, the accretion flow eventually migrates inward to the radius at which its angular momentum imposes itself, and the flow eventually circularizes and forms a disk. That disk, however, is likely to be thick, inhomogeneous, and time-variable in its orientation, given the variability of the locations and the angular momenta of the sources of the accreting material.

Submillimeter observations of the H30$\alpha$ hydrogen recombination line have been reported arising from the inner accretion flow onto Sgr A* by Murchikova et al. (2019; see also Yusef-Zadeh et al. 2020); they interpret their very broad (2200 km s$^{-1}$), double-peaked line profile and a positional displacement of ±0.11" (±0.005 pc) between the blue and red sides of the profile in terms of emission from an accretion disk. Except for X-ray lines from highly ionized material much further out near the Bondi radius (1.5", Wang et al. 2013), the H30$\alpha$ line is the only spectral line so far seen from the accretion flow. Indeed, the hydrogen Brackett-$\gamma$ line is absent at such a low level (Ciurlo et al. 2021) that it led Murchikova et al. (2019) to argue that the H30$\alpha$ line must be amplified by a factor of 80 via maser amplification.[5] The formation of a "disk-like structure" of radius ~0.01 pc at temperatures favoring recombination line emission (~10$^4$ K) emerges from the hydrodynamical models of Calderón et al. (2020), based on accretion from the surrounding stellar winds. However, the disk mass in their models (~5 x 10$^{-3}$ M$_\odot$) is one or two orders of magnitude larger than the disk mass estimated by Murchikova et al. (2019).

---

[5] Because Ciurlo et al. (2021) question this important result, it clearly warrants verification.

Other models (Ressler et al. 2020a,b) indicate that circularization doesn't happen except on much smaller scales where the concentrated magnetic field becomes strong enough to assert itself, forming a *magnetically arrested disk* with a radius of a few tens of gravitational radii. This is because the infall velocities at larger distances are comparable to the free-fall time, and because the radiative cooling efficiency is low on such timescales, so the infalling gas heats up and has a tendency to expand back outwards.

The low luminosity of Sgr A* implies that the radiative efficiency of the material that does accrete onto the GBH is very low, as many authors have noted, and have discussed in terms of various categories of a radiatively inefficient accretion flow (Melia & Falcke 2001, and references therein). The accretion flow onto the GBH has been interpreted as a hot, optically thin accretion disk, heated by viscous dissipation of the accretion energy. In this plasma, the ion temperature is much higher than the electron temperature because of the much higher radiative efficiency of the electrons (Shapiro et al. 1976). The thermal energy is then advected inward by the accretion flow, giving rise to the generic term Advection-Dominated Accretion Flow, or ADAF (Narayan & Yi 1994; 1995; Yuan and Narayan 2014, and references therein). The ions stay hotter than the electrons because the thermal equilibration time scale far exceeds the accretion time.

Two principal variants of the ADAF model have been forwarded: the convection-dominated accretion flow (CDAF) and the Adiabatic Inflow-Outflow Solution (ADIOS). The CDAF model is based on the assumption that hot accretion flows are convectively unstable (Narayan, Igumenshchev & Abramowicz 2000; Quataert & Gruzinov 2000). The accreting fluid therefore becomes stalled in convective eddies, with the consequence that there is an inward decrease of the accretion rate. The ADIOS model posits that much of the accretion energy acquired by the disk drives a wind from the disk, so there, too, the accretion rate decreases inward (Blandford & Begelman 1999; 2004; Shcherbakov & Baganoff 2010; Begelman 2012; Yuan et al. 2012a, b).

## 6. Outflows from the GBH

6.1 An Outflowing Wind from the GBH Accretion Flow

The vast majority of the gas falling through the Bondi radius does not make it to the event horizon. Measurements of the Faraday rotation of the millimeter and submillimeter synchrotron emission from Sgr A* indicate that the accretion rate must be less than or on the order of $1 - 3 \times 10^{-8}$ M$_\odot$ yr$^{-1}$ (Bower et al. 2003; Marrone et al. 2007). Such a low rate is borne out by radiatively inefficient accretion models and MHD models of geometrically thick accretion flows (Sharma et al. 2007a, b; Ressler et al. 2020b). As a consequence of the fact that the accretion rate onto the GBH is about 3 orders of magnitude smaller than the rate of material passing through the Bondi radius, there must be an outflowing wind, presumably somewhat collimated by the geometry of the accretion flow. Most of the outflow arises well outside the innermost regions, at distances exceeding ~10 gravitational radii (Yuan et al. 2012a).

Observationally, there is ample evidence for an outflow, perhaps quite variable or intermittent, from the GBH. Radio images, for example, show features such as bow shocks that can be attributed to an outflow (Yusef-Zadeh et al. 2015; Zhao et al. 2016). Infrared images also show bow-shock sources near the GBH (Muzic et al. 2010). However, in both the radio and infrared, the bow-shock sources could have resulted from the collective winds of the WR stars in the young nuclear cluster, so that evidence for a wind from the GBH is not conclusive. Another radio feature for which a collimated wind from Sgr A* has been suggested is a pair of nonthermal radio structures located 5 - 10 pc away on opposite sides of the GBH which are nonthermally emitting structures showing very strongly curved magnetic filaments (Morris, Zhao & Goss 2014). The center of curvature of these distorted filaments (named the northern and southern curls) is toward the GBH, consistent with the possibility that a collimated wind, or even a broad jet, propagated out in that direction in the recent past.

X-ray images also show features that can be attributed to collimated outflows from Sgr A*. On a scale of ~ ± 15 pc, there is a bipolar X-ray source straddling the location of the GBH and oriented perpendicular to the Galactic plane (Morris et al. 2003; Ponti et al. 2015a). The bipolar feature could have resulted from a large outburst at the GBH within the past $10^4$ years – a tidal disruption event of a star, for example, although a supernova remnant channeled by a strong vertical magnetic field cannot be ruled out as the cause of the bipolar structure.

On scales of several hundred parsecs, extended X-ray and radio data show the presence of "chimneys" – quasi-cylindrical emission structures extending to both sides of the Galactic plane (Ponti et al. 2019, 2021; Nakashima et al. 2019; Heywood et al. 2019). X-ray emission arises from the interior of the chimneys, heralding a hot outflow arising from somewhere in the central ~100 pc, while radio emission arises from the periphery of the chimneys, presumably from an interface between their hot interiors and the ambient interstellar medium. Episodic ejections from the GBH caused by occasional large accretion events or tidal disruptions of stars have almost certainly contributed to the sculpting of the chimneys, but a quasi-continuous wind from the GBH could also be an important contributor. Again, however, relatively frequent supernovae occurring in the central ~100 pc could also be an important or even dominant contributor to the fabrication of the chimneys (Zhang, Li & Morris 2021).

On the largest scales (up to ± 15 kpc), the γ-ray emitting Fermi Bubbles (e.g., Yang et al. 2018, 2022) and the even larger eRosita X-ray Bubbles (Predehl et al. 2020) are other giant forms of bipolar nebulae that have possibly been energized by high-velocity emanations from the GBH. The Fermi Bubbles could have been created by powerful collimated outflows in a single AGN-level event several million years ago (Zubovas, King & Nayakshin 2011; Zubovas & Nayakshin 2012), or by winds generated in a continuous series of tidal disruption events by the GBH extending over a much longer period of time (Ko et al. 2020). The energetic particles and hot plasma that are characteristic of the Fermi Bubbles have likely been fed through the X-ray chimneys discussed above (Ponti et al. 2019). As previously stated, there are potential supernova alternatives for powering the outflows that have generated the Fermi bubbles. A model in which star formation in the central molecular zone leads to supernova-produced cosmic rays that diffuse upward into the Fermi Bubbles, leading to hadronically-produced, γ-rays has been proposed by Crocker & Aharonian (2011), Crocker (2012), and Crocker et al. (2015). Further survey work is presently under way with both the Chandra and XMM-Newton X-ray observatories to elucidate the roles of the GBH versus star formation in generating the Fermi Bubbles.

6.2  Does the GBH Produce a Jet?

Accreting black holes of all masses produce jets when the accretion flow is sufficiently organized to form a well-defined disk, or when the accreting material carries magnetic flux into a black hole having a substantial spin, in which case the Blandford-Znajec (1977) mechanism operates to drive Poynting flux out along the spin axis that can accelerate particles into a collimated beam.  The degree of collimation depends on the geometry and the inertia of the collimating accretion disk as well as on the black hole's spin.  In the case of the GBH, no *radio* jet has yet been directly observed, although a considerable amount of work has been done to interpret the variable radio emission from Sgr A* in terms of emission from a jet (Falcke et al. 1993; Falcke & Markoff 2000; Markoff et al. 2001; Yuan, Markoff & Falcke 2002; Markoff, Bower & Falcke 2007; Falcke, Markoff & Bower 2009; Moscibrodzka & Falcke 2013).

Various claims have been made for observable, large-scale morphological manifestations of a jet from Sgr A* (Yusef-Zadeh et al. 1986; 2012, 2016, 2020; Sofue et al. 1989; Su & Finkbeiner 2012; Li et al. 2013).  The jet candidate discussed by Li et al. (2013, and originally considered as a jet by Morris et al. 2004 and Muno et al. 2008) is based on a ~1 pc linear X-ray feature that is aligned with the location of Sgr A*, but starts about 1 pc away at the location of a shock front (Figure 11).  Li et al. (2013) argued that the shock front is created where a broadly collimated wind centered on the X-ray jet impacts an orbiting gaseous stream known as the Eastern Arm of the Sgr A West HII region.  The shock creates a discontinuity in the velocity field.   Follow-up work by Zhu et al. (2019) with 5.6 megaseconds of X-ray data from the Chandra X-ray Observatory spanning 18 years shows that the spectrum of the linear feature is unusually hard at the location closest to Sgr A* and that it undergoes significant spectral softening with increasing distance from Sgr A*.  The hard spectrum is Interpreted as resulting from rapid cooling of the jet electrons in the immediate post-shock region, leading to a piling up in energy, thus creating a quasi-monoenergetic distribution that cools by synchrotron radiation along the jet path.  The X-ray data also show that the jet luminosity is invariant over 18 years, with the possible exception of the measurements in 2002, about one year after the extended, dusty object G1 orbited through its periapse at a distance of only about 300 AU from the GBH, where it apparently had some material tidally removed (Witzel et al. 2017).  The newly unbound material presumably joined the accretion flow onto the GBH and might thereby have been responsible for powering a factor-of-2 increase in the X-ray luminosity of the putative jet in 2002, which is significant at the 2-σ level.

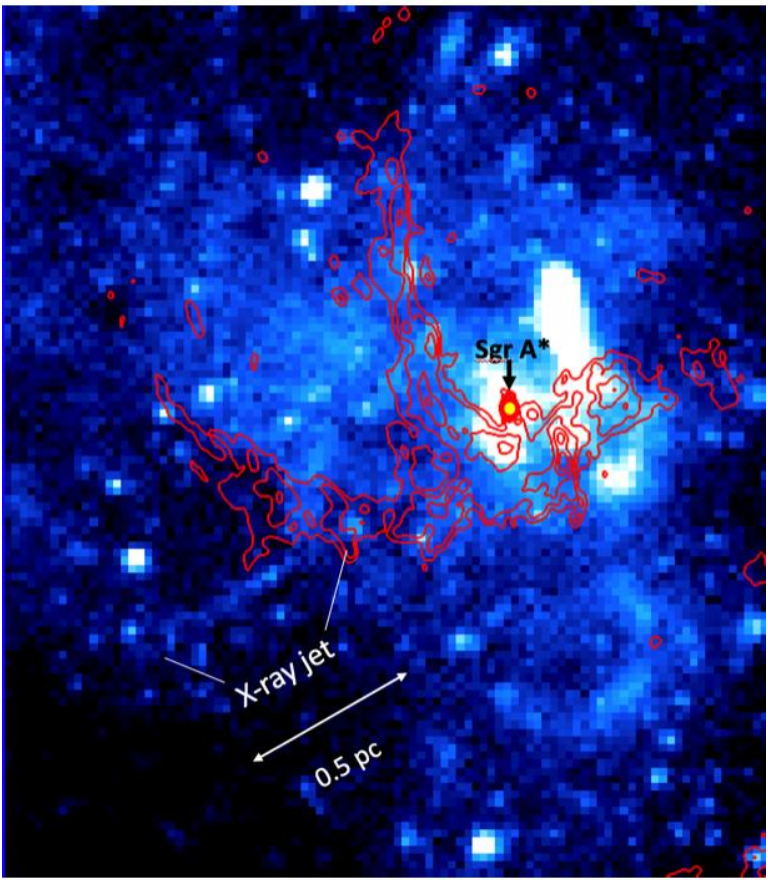

**Figure 11:** 2 - 8 keV X-ray image of the central few parsecs of the Galaxy, showing the location of the hypothetical X-ray jet from Sgr A* (data from Zhu et al. 2019). Contours of 6-cm radio continuum emission from Zhao et al. (2016) are superimposed.  In this image, equatorial north is up.  The jet feature is aligned with the Galaxy's rotation axis.  The observable portion of the jet begins at the location of a shock that appears here as an enhanced radio contour.

## 7.  New Perspectives and Future Prospects

Sagittarius A* has been a prime target of the Event Horizon Telescope (EHT), which has recently provided exciting new images of the region immediately surrounding the event horizon of the GBH (Figure 12, from EHT Collaboration 2022).  The image shows a bright ring surrounding the black hole "shadow", and the diameter of the ring, 51.8 ± 2.3 µas (0.42 AU), is consistent with that expected from general relativistic models of the accretion flow onto a 4 million solar mass Kerr black hole at the distance of the Galactic center.

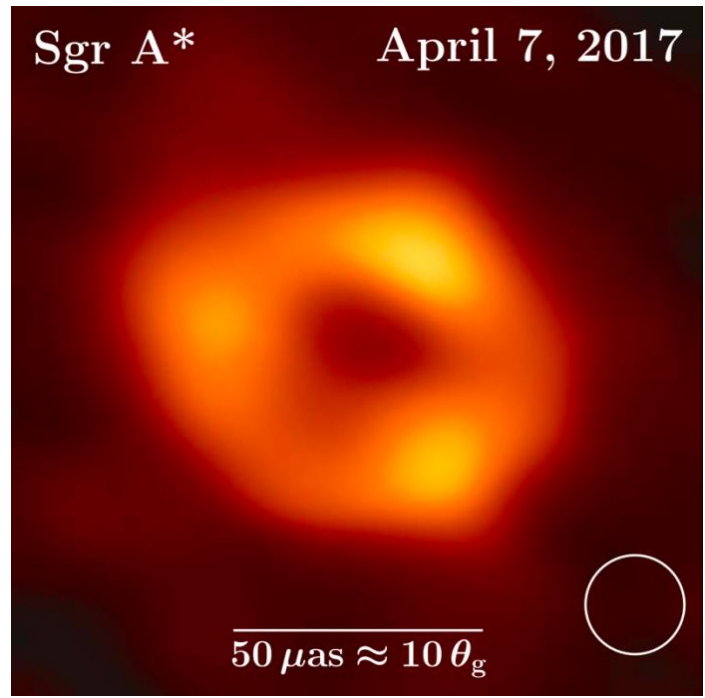

**Figure 12**: Representative EHT image of Sgr A* at a wavelength of 1.3 mm (Event Horizon Telescope Collaboration 2022a). This image was made by averaging the images produced by different reconstruction methodologies and reconstructed morphologies.   The circle at lower right shows the restoring beam used for CLEAN image reconstructions (20 µas FWHM).  The peak brightness temperature is about 1.2 x $10^9$ K.

Since Sgr A* varies on timescales as short as minutes (Witzel et al. 2021; Wielgus et al. 2022a), the challenge presented to a very-long-baseline interferometer such as the EHT has been to capture a sufficient sampling of the UV plane, i.e., a sufficient coverage of projected baselines, in the brief time interval over which the source undergoes significant variations. Rather than being able to assume static images as was done for the much slower-varying black hole in M87 (EHT Collaboration 2019), the EHT collaboration has employed various statistical strategies for producing a time-averaged image (EHT Collaboration 2022b).

EHT data on Sgr A* also place constraints on the magnitude and orientation of the black hole spin. The EHT Collaboration concludes that the spin is non-zero and prograde, and that the preferred inclination of the spin axis is $\leq 50^{\circ}$. This means that the equatorial plane is unlikely to be edge-on, so that the orientation of the spin axis of the GBH is very different from that of the Galaxy. The EHT has already acquired, and will continue to acquire, more data on Sgr A* so one can expect a refinement of these initial conclusions as more data are obtained and as the capabilities of the EHT evolve with an increasing number of stations.

On another front, the next generation of large optical/IR telescopes will provide much greater precision on stellar orbits as well as access to fainter stars and gaseous features on more tightly bound orbits around the GBH. That will lead to more stringent tests of general relativity, and when coupled with continued EHT observations, we can expect future work to bring exciting information on the accretion process near the event horizon of a spinning black hole, and perhaps on the launching of a jet.

Finally, finding a closely orbiting pulsar could provide a revolutionary testbed for GR and for measuring the GBH characteristics (Pfahl & Loeb 2004), and dedicated but unsuccessful searches have been made for a sufficiently close-in pulsar (e.g., Eatough et al. 2021), but with anticipated near-future facilities such as the Square Kilometer Array and the Next Generation VLA, deeper searches will certainly be warranted.